\documentclass{aa}
\usepackage{graphicx,color,subfigure,multirow}
\usepackage{txfonts}
\usepackage{verbatim}
\graphicspath{{./figures/}}
\usepackage{natbib,twoopt}
\usepackage[breaklinks=true]{hyperref}
\hypersetup{
        colorlinks=true, 
        urlcolor=blue,
        linkcolor=blue
}
\bibpunct{(}{)}{;}{a}{}{,}

\makeatletter
\newcommand{\bibnote}[2]{\global\@namedef{#1note}{#2}}
\newcommand{\biblink}[2]{\global\@namedef{#1link}{#2}}

\makeatother
\makeatletter
\protected\def\stonyslink{%
        \def\hyper@linkstart##1##2{}\let\hyper@linkend\@empty}
\newcommandtwoopt{\citeads}[3][][]{%
        \href{http://adsabs.harvard.edu/abs/#3}%
        {\stonyslink \citealp[#1][#2]{#3}}%
        \biblink{#3}{\href{http://adsabs.harvard.edu/abs/#3}{ADS}}}
\newcommandtwoopt{\citepads}[3][][]{%
        \href{http://adsabs.harvard.edu/abs/#3}%
        {\stonyslink \citep[#1][#2]{#3}}%
        \biblink{#3}{\href{http://adsabs.harvard.edu/abs/#3}{ADS}}}
\newcommandtwoopt{\citetads}[3][][]{%
        \href{http://adsabs.harvard.edu/abs/#3}%
        {\stonyslink \citet[#1][#2]{#3}}%
        \biblink{#3}{\href{http://adsabs.harvard.edu/abs/#3}{ADS}}}
\newcommandtwoopt{\citeyearads}[3][][]{%
        \href{http://adsabs.harvard.edu/abs/#3}%
        {\stonyslink \citeyear[#1][#2]{#3}}%
        \biblink{#3}{\href{http://adsabs.harvard.edu/abs/#3}{ADS}}}
\makeatother

\begin{document}

\title{Numerical estimation of the capture
ability of Neptunian mean motion resonances}
\author{
    Hailiang Li\inst{1}
    \and
        Li-Yong Zhou\inst{2}\fnmsep\inst{3}
        \and
    Xiaoping Zhang\inst{1}
        }
\authorrunning{Li et al.}
\institute{State Key Laboratory of Lunar and Planetary Sciences, Macau University of Science and Technology, Macau 999078, China\\
            \href{mailto:xpzhang@must.edu.mo}{xpzhang@must.edu.mo}
            \and
            School of Astronomy and Space Science, Nanjing University, 163 Xianlin Avenue, Nanjing 210046, China
              \and
            Key Laboratory of Modern Astronomy and Astrophysics in Ministry of Education, Nanjing University, China
        }
\date{}

\abstract{

Resonant populations of trans-Neptunian objects serve as crucial dynamical archives for unraveling the early migratory history of the Solar System. A quantitative assessment of the capture efficiency into various mean motion resonances (MMRs) during migration is essential for understanding the origins of these populations, constraining migration parameters, and reconstructing of the primordial planetesimal disk. Using numerical simulations, this study systematically investigates the capture capability of exterior MMRs during Neptune's outward migration in a planar model. For a specific $p$:$q$ MMR, the small bodies can be captured only when their eccentricities surpass a certain threshold, $e_{min}$, which increases with faster migration rates, greater distances of MMRs, and higher resonance orders. We also find that 1:$q$-type MMRs exhibit notably higher $e_{min}$ due to their unique dynamical structure. On the other hand, as long as a particle's eccentricity is suitable, its capture efficiency shows little dependence on the migration rate; instead, it mainly depends on the $p$ value and heliocentric distance, decaying exponentially as either parameter increases. Based on our simulation results, we derive for the first time a simple empirical expression to calculate $e_{min}$ and the capture efficiency. From beyond the 1:2 MMR out to approximately the 1:4 MMR, the theoretically predicted capture numbers follow a trend that resembles what is seen in observations, suggesting that migration capture could be the primary source of resonant populations in these regions. However, in more distant regions, the theoretical predictions fall significantly short of observational estimates, implying that other mechanisms (e.g., resonant sticking) might be necessary. This research provides a systematic quantitative framework for understanding capture into Neptunian MMRs during migration. Future integrations of more comprehensive observational data will facilitate a more precise reconstruction of the Solar System's early dynamical evolution.
}

\keywords{celestial mechanics -- Kuiper belt: general -- methods: numerical
}
\maketitle{}

\section{Introduction}

Within the context of planetary migration, the distribution patterns of the vast majority of trans-Neptunian objects (TNOs) can now be explained \citepads[e.g.,][]{Levison2008, Morbidelli2020}. Among these, Neptune's mean motion resonances (MMRs) contain significant information about the early Solar System. As more TNOs are discovered, the insights embedded within Neptune's MMRs, particularly those related to planetary migration, can be further elucidated.

Many mechanisms can influence both the number and distribution of small bodies within resonance belts. For instance, numerous studies have compared the stability of different MMRs \citepads[e.g.,][]{Morbidelli1995, Melita2000, Tiscareno2009, Saillenfest2016, Li2024}. This stability reflects the likelihood of objects escaping from MMRs after being captured. Other fundamental factors include the capture ability and capture locations of MMRs, which result in the distinct population sizes retained by different MMRs since the early Solar System.

A substantial  number of prior studies have  investigated the physical process of resonant capture. Early work often treated it as an adiabatic process, which required migration to be sufficiently slow, with a timescale that is significantly longer than the libration period of the resonant angle. Analytical methods have derived capture probabilities under idealized models \citepads[e.g.,][]{Peale1976, Yoder1979, Henrard1982, Borderies1984, Murray1999}. A key conclusion of adiabatic capture theory is that when migration is sufficiently slow and the particle's eccentricity is below a certain threshold, capture is inevitable. Conversely, if the particle's eccentricity exceeds this threshold, the capture probability decreases progressively \citepads{Borderies1984}.

\citetads{Wyatt2003} further extended the study to higher migration rates, focusing on several specific MMRs and assuming a dynamically cold planetesimal disk. This work demonstrated that the capture probability increases with planetary mass and decreases with migration rates. Building on this, subsequent work employed Hamiltonian models to conduct a detailed exploration of first- and second-order MMRs, further investigating how different initial eccentricities of small bodies affect capture efficiency \citepads{Quillen2006, Mustill2011}. These studies also systematically examined issues such as the interference of nearby resonant terms, the shifting of resonant centers during migration, the changes in eccentricity of particles upon failed capture, and the amplitude distribution of captured particles.

With advancements in technology and observations, resonant TNOs have been discovered in dozens of Neptunian MMRs, extending out to hundreds of astronomical units. Several studies have provided estimates of the number of objects within these resonant belts \citepads[e.g.,][]{Gladman2012, Volk2016, Crompvoets2022}. Through direct numerical simulations within a planar model, we aim to test the ability of different Neptunian MMRs to capture bodies, quantify key parameters (e.g., the eccentricity threshold and probability for capture), and extend this analysis to higher order MMRs. Furthermore, we compare the predicted number of captured bodies in each MMR from our model with observational resonant populations, aiming to constrain the early dynamical evolution of the Solar System.

The remaining sections of this paper are organized as follows. In Section 2, we describe the numerical methods, the initial conditions, and the approach used to identify resonant objects. In Section 3, we analyze the simulation results, evaluate the capture capabilities of MMRs between 30\,au to 80\,au, carry out a detailed individual analysis of 1:$q$ MMRs, and provide a comparison with observational estimates. Finally, we conclude with a summary and discussion.

\section{Methods}

\subsection{Initial conditions}
In this study, we employed a planar restricted three-body model consisting of the Sun, Neptune, and massless small bodies. Although other giant planets may significantly alter Neptune's migration \citepads[e.g.,][]{Tsiganis2005, Nesvorny2012} and continue to affect the stability of MMRs after migration ends \citepads[e.g.,][]{Li2024, Graham2024}, this paper focuses solely on the physical process of capturing small bodies, which is predominantly governed by Neptune itself.

For a given $p$:$q$ MMR (where $p$<$q$), its strength primarily depends on the resonance order $k=q-p$, with lower order MMRs generally exhibiting a greater strength. The nominal location of the MMR is given by $a_N\times(\frac{q}{p})^{2/3}$, where $a_N$ denotes Neptune's semimajor axis. Under the simplified model considered in this paper, the resonant angle for an MMR can be expressed as $\phi=q\lambda-p\lambda_N-(q-p)\varpi$, where $\lambda$ and $\lambda_N$, respectively, represent the mean longitude of the small body and Neptune, while $\varpi$ represents the longitude of perihelion of the small body.

In each simulation, we applied an additional force in the direction of Neptune's velocity, making Neptune migrate at a constant rate of $\dot{a}_N$. During this process, its eccentricity remains small and it can be considered on a circular orbit. Numerous previous studies have examined the influence of Neptune's migration on resonance capture \citepads[e.g.,][]{Chiang2002, Li2014, Kaib2016, Li2023}. While there is no consensus yet on the precise details of Neptune's migration, it is generally believed that during the early Solar System, a phase known as the giant planet instability occurred, during which the giant planets underwent dramatic orbital changes \citepads[see][for a review]{Nesvorny2018}. Therefore, we explored a wide range of $\dot{a}_N$ values (from 0.01\,au/Myr to 10\,au/Myr), allowing each set of simulations to analogously represent MMR capture scenarios at different migration stages. In each simulation, the $\dot{a}_N$ remains constant.
 
We first conducted a general statistical analysis of capture. In this setup, small bodies were distributed between 30\,au and 80\,au, with randomized semimajor axes. In other words, the surface density of the planetesimal disk is proportional to $r^{-1}$, where $r$ represents the heliocentric distance. The total number of particles is 50,000. In these simulations, Neptune migrates from 27\,au to 30\,au. The initial eccentricity $e_0$ of the particles is randomly distributed between 0 and 0.35, while the angular elements, $\varpi$ and $M$, are randomly distributed between $0^\circ$ and $360^\circ$. For each MMR, we recorded the minimum initial eccentricity of the captured objects as $e_{min}$ and the number of objects captured into MMR as $P_{res}$.

Since we observed that 1:$q$ MMRs required a higher $e_{min}$ to capture objects compared to other MMRs, we conducted an additional group of simulations focusing on this specific category. In these simulations, Neptune migrates from 29\,au to 30\,au and the semimajor axes of the planetesimals are set to the nominal resonance location corresponding to Neptune at 29.5\,au (i.e., $29.5\times q^{2/3}$). In this simulation set, 180 $\varpi$ values were sampled uniformly from $0^\circ$ to $358^\circ$ and $M$ was fixed at $0^\circ$. Additionally, we tested multiple sets of different $e_0$ until the value of $e_{min}$ could be determined within a precision of one-thousandth.

The integrations in this study are performed using the N-body code REBOUND \citepads{rebound}, employing the WHFast algorithm, a Wisdom–Holman symplectic integrator \citepads{reboundwhfast, wh}. Depending on the values of $\dot{a}_N$ and the total migration distance, the duration of Neptune's migration varied, while in all cases it stops upon reaching 30\,au. After this phase, the additional force driving migration is removed and each simulation is extended for another 0.5\,Myr, with relatively frequent output intervals to determine whether each object is within the MMR.

\subsection{Resonator identification}

\begin{figure}[!htb]
\centering
\resizebox{\hsize}{!}{\includegraphics{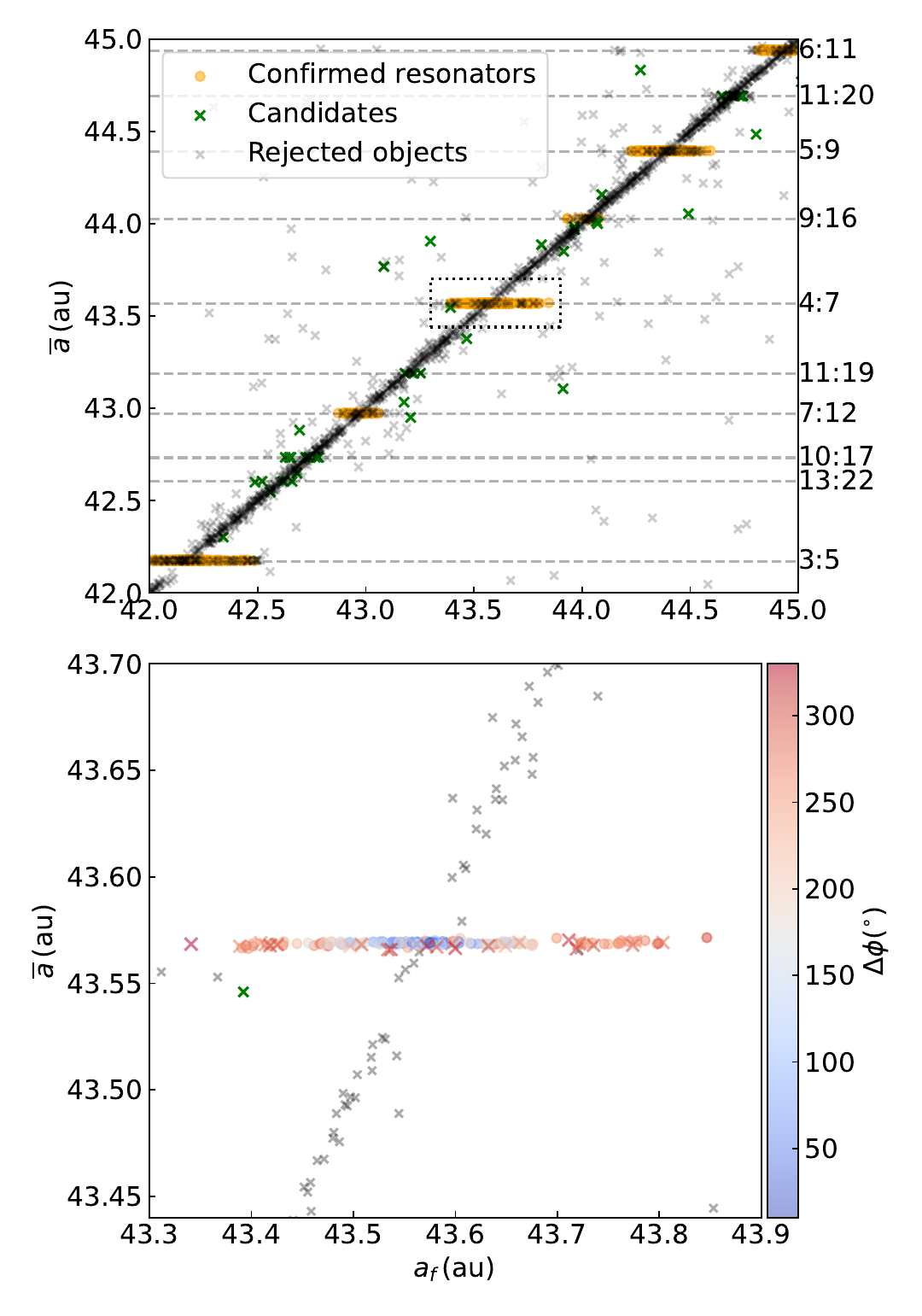}}
\caption{Resonance identification procedure based on semimajor axis. The horizontal axis represents the final semimajor axis, $a_f$, of the particles, while the vertical axis represents their mean semimajor axis, $\overline{a}$, during the final 0.5 Myr. The figure shows the region near the 4:7 MMR in the simulation with $\dot{a}_N=0.2$\,au/Myr. In the upper panel, black crosses indicate objects excluded during the initial filtering, green crosses represent resonant candidates that were subsequently excluded during clustering, and orange dots denote particles finally identified as resonators. Horizontal dashed lines mark the positions of several MMRs. The lower panel provides a magnified view of the dotted rectangular region in the upper panel, where all particles residing in the 4:7 MMR are identified based on the libration of their resonant angles, with colors indicating the corresponding libration amplitudes.}
\label{fig:Res_identify}
\end{figure}

Following the simulation setup described in the previous subsection, we first determined whether the particles reside within MMRs. Traditionally, this would require assessing whether their resonant angle is in libration. However, in this simulation, we did not know in advance whether a given particle was in MMR or which specific MMR it might occupy. Therefore, we adopted a semimajor axis-based method to make this determination. To this end, we denoted the particle's initial and final semimajor axis as $a_i$ and $a_f$, respectively. The mean semimajor axis during the final 0.5 Myr of the simulation, when Neptune no longer migrates, was recorded as $\overline{a}$. In Fig.~\ref{fig:Res_identify}, we illustrate the process of identifying resonant objects, using the simulation with $\dot{a}_N=0.2$\,au/Myr as an example. As shown in the upper panel of Fig.~\ref{fig:Res_identify}, the $\overline{a}$ value of small bodies exhibits clear clustering at MMR locations, which are marked by horizontal dashed lines. This clustering  served as the key basis for identifying resonators.

If a particle is captured by an MMR and migrates outward with Neptune, its orbit also expands accordingly. Based on their $a_i$ and $\overline{a}$, we first filtered particles captured by Neptune during the first 90\% of its migration path. This requires the particle to have been captured before Neptune reaches 29.7\,au, namely, satisfying $\frac{\overline{a}}{a_i}>\frac{30}{29.7}$. Additionally, we excluded particles with eccentricities greater than 0.45, as well as those whose semimajor axis varied by more than 2\,au during the final 0.5\,Myr.

This filter excludes the following three types of particles: (1) particles that experience minimal resonant influence; their semimajor axes remain almost unchanged or slightly reduced, resulting from  MMRs crossing without being captured \citepads{Mustill2011}. These particles generally have relatively low eccentricities and are distributed approximately along the diagonal from the lower left to the upper right in Fig.~\ref{fig:Res_identify}; (2) particles captured during the final 10\% of Neptune's migration path. In our model, Neptune migrates outward at a constant rate and stops abruptly upon reaching 30\,au. As noted by \citetads{Li2023}, in such cases, the final capture scenarios are different from those during the migration, due to the deceleration or stopping of migration. Therefore, we excluded those particles when statistically analyzing resonant objects. This filtering step treats different MMRs fairly and does not impact the comparisons between them. This category of particles, as with other resonant bodies, is also located near the MMR positions shown in Fig.~\ref{fig:Res_identify}; and (3) objects that primarily belong to the scattered disk and usually have a lower perihelion distance. Due to their significant variations in semimajor axis during the final 0.5\,Myr, their positions are primarily located far from the diagonal region in Fig.~\ref{fig:Res_identify}.

Since resonant objects oscillate around the nominal resonance location, their $\overline{a}$ are very close to the nominal value, typically with a deviation of less than 0.01\,au, as illustrated in the lower panel of Fig.~\ref{fig:Res_identify}. Accordingly, we group particles based on their $\overline{a}$, with each group representing an MMR or, in a few cases, isolated particles not captured into any MMRs. We further excluded groups with less than 10 particles, ensuring that nonresonant particles were filtered out and eliminating MMRs with few members. In the example shown in Fig.~\ref{fig:Res_identify}, several high-order MMRs (e.g., 13:22, 10:17, 11:19, and 11:20) were excluded due to an insufficient number of particles. This step is reasonable as we need to statistically assess $e_{min}$ and $P_{res}$ for each MMR; groups with too few particles would lack statistical significance, and the derived $e_{min}$ could deviate substantially from the true value.

To validate the accuracy of this resonator identification method, we also checked the resonant angles of all small bodies near the 4:7 MMR. Objects with librating resonant angles are shown in the lower panel of Fig.~\ref{fig:ae_final}, with colors indicating their libration amplitudes of resonant angle. We found that all small bodies selected by the aforementioned method exhibit librating resonance angles, confirming the effectiveness of the approach. For some objects, while their resonant angles were librating, they were excluded due to insufficient semimajor axis increment. These objects were captured during the final 10\% of Neptune's migration and located at the edge of the resonance island, resulting in larger libration amplitudes.

\section{Results}

\begin{figure}[!htb]
\centering
\resizebox{\hsize}{!}{\includegraphics{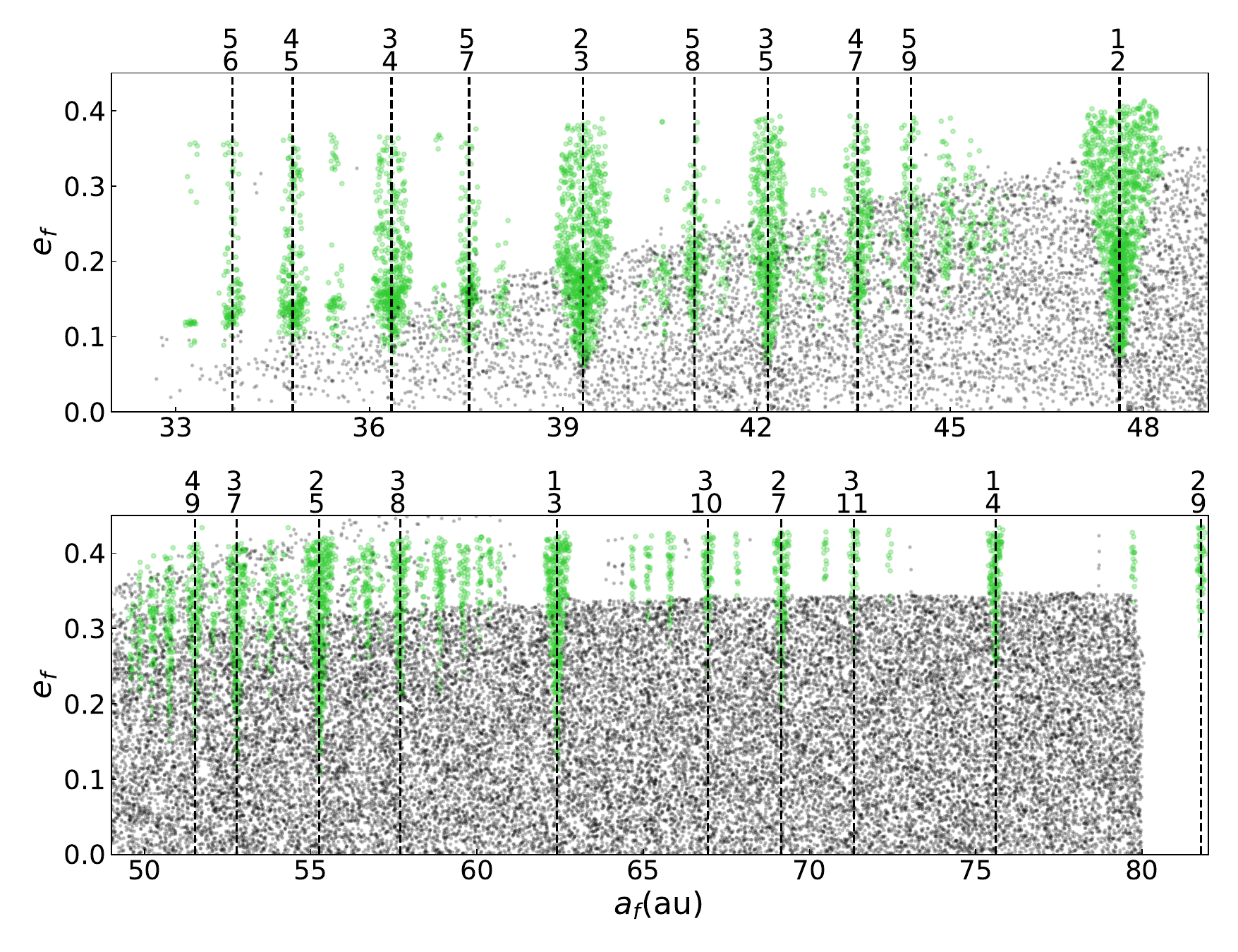}}
\caption{Distribution of final semimajor axis, $a_f$, and final eccentricity, $e_f$, from simulations with $\dot{a}_N=0.2$\,au/Myr. Green points represent objects identified as being in MMRs, while the gray points indicate the others. The positions of some major MMRs are marked with dashed lines, with the corresponding $p$ (upper number) and $q$ (lower number) values indicated above the lines. For clarity, the figure is divided into two panels using 49\,au as the boundary, showing the inner and outer regions separately.}
\label{fig:ae_final}
\end{figure}

In this section, we present the simulation results and our understanding the capturing capacity of each MMR. This not only validates previous theories of resonant capture \citepads[][]{Borderies1984, Wyatt2003, Quillen2006, Mustill2011}, but also extends these theories to higher order MMRs. As an example, Fig.~\ref{fig:ae_final} illustrates the simulation results when $\dot{a}_N=0.2$\,au/Myr.

In Fig.~\ref{fig:ae_final}, the vertical band-like structures in green can be observed, each corresponding to an MMR. It is evident that the majority of resonant objects are concentrated in the inner region. Within 50\,au, even the ones that have not been captured by MMR have had their orbits significantly modified, either by Neptune's scattering or MMRs' sweeping. Beyond 60\,au, the influential MMRs become relatively sparser and the number of resonant objects decreases significantly. Meanwhile, particles that were not captured largely remained near their initial positions and experienced limited influence from the MMR sweeping. Furthermore, Fig.~\ref{fig:ae_final} illustrates that stronger resonances exhibit broader resonance widths and are able to capture small bodies at lower eccentricities, a phenomenon we discussed in detail later in this paper.

\begin{figure*}[!htb]
\centering
\resizebox{\hsize}{!}{\includegraphics{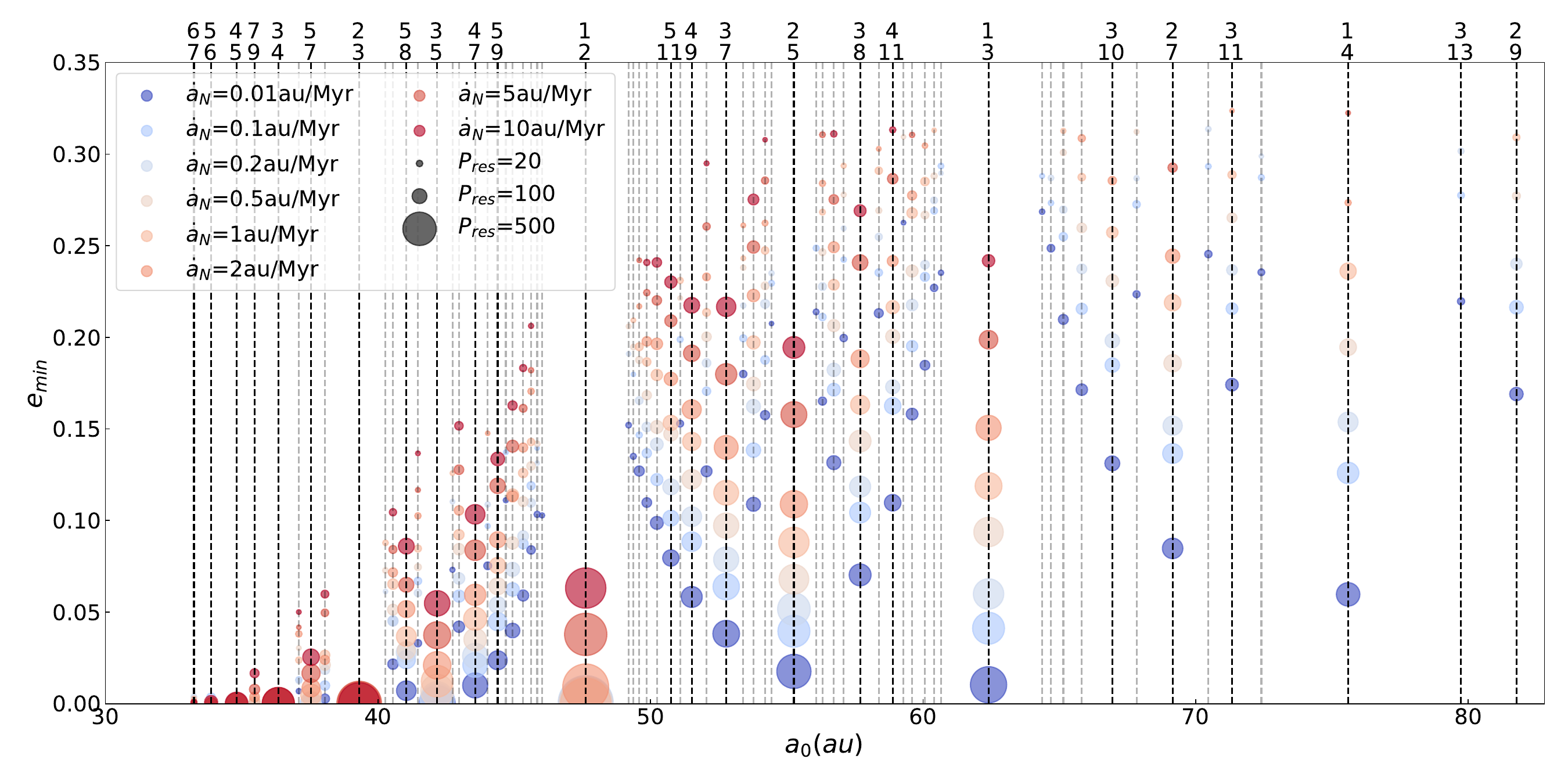}}
\caption{ $e_{min}$ and $P_{res}$ for different MMRs under various $\dot{a}_N$ values. The horizontal axis represents the final location of each MMR, the vertical axis shows the minimum initial eccentricity, $e_{min}$, for each MMR, the color of the points indicates Neptune’s migration rate $\dot{a}_N$ in the simulation, and the size of the points represents the number of captured small bodies, $P_{res}$. Some of the major MMRs are marked with dashed black lines, with their corresponding $p$ and $q$ values indicated above the lines, while other MMRs are shown with gray dashed lines.}
\label{fig:e_minimum}
\end{figure*}

We conducted eight simulations with different $\dot{a}_N$ values and observed 74 MMRs that captured more than ten particles in at least one simulation. Among the outermost MMRs, we retained the 2:9 MMR located near 81.8\,au (visible on the far right of Fig.~\ref{fig:ae_final}), as its swept area predominantly lies within 80\,au. The 3:14 MMR and the 1:5 MMR were excluded, despite also capturing a number of objects. To minimize the influence of randomness, we selected 66 MMRs that captured more than 10 particles in at least two simulations. The highest order MMR among these is the 9:25 MMR at 59.3\,au, whose order is 16. For these MMRs, we computed their $e_{min}$ and $P_{res}$,  summarizing the results in Fig.~\ref{fig:e_minimum}.

In most cases, capture occurs only when the eccentricity of the small body exceeds a certain threshold, corroborating previous findings by \citetads[e.g.][]{Mustill2011}. Some patterns can be directly observed in Fig.~\ref{fig:e_minimum}. For example, higher order MMRs generally require a higher $e_{min}$ compared to lower order MMRs, and a faster $\dot{a}_N$ also raises this threshold. On the other hand, the $P_{res}$ varies significantly among different MMRs, while is relatively insensitive to the $\dot{a}_N$. Nevertheless, for certain MMRs (e.g., the 1:3 MMR), $P_{res}$ decreases notably as $\dot{a}_N$ increases. The main reason is that the $e_{min}$ increases with faster $\dot{a}_N$. As a result, the number of particles eligible for capture correspondingly decreases. In the following, we discuss the variations of $e_{min}$ and $P_{res}$ across different MMRs and a range of $\dot{a}_N$.

\subsection{Minimum eccentricity for MMR capture}

To discuss the distribution of $e_{min}$ across different $\dot{a}_N$ and MMRs, we first need to identify those MMRs capable of capturing small bodies even when they are in circular orbits. Therefore, we defined that if a MMR has an $e_{min}$ smaller than 0.005, it is considered capable of capturing objects at $e_0=0$. The method used to initialize the small bodies ensures that there are approximately 1,000 small bodies per astronomical unit, of which about 14 have $e_0<0.005$. Since Neptune migrates over a distance of 3\,au, an exterior MMR necessarily sweeps through a distance greater than 3\,au. Therefore, each MMR would have encountered at least several dozen objects with $e_0<0.005$.

First, we focus on the first-order MMRs separately, as they typically exhibit efficient capture down to zero initial eccentricity. Our statistics included six first-order MMRs, from the 1:2 MMR to the 6:7 MMR, as shown in Fig.~\ref{fig:e_minimum}. The 1:2 MMR can capture objects with $e_0=0$ only when $\dot{a}_N \leq 1$\,au/Myr, whereas other first-order MMRs can do so under any $\dot{a}_N$ up to $10$\,au/Myr. This is likely related to the fact that first-order MMRs retain a certain degree of resonance width even as eccentricity approaches zero \citepads[e.g.,][]{Murray1999,Morbidelli2002}. In addition to these six first-order MMRs, three second-order MMRs and one third-order MMR are also capable of capturing particles with $e_0=0$ under sufficiently slow migration: the 3:5 MMR ($\dot{a}_N \leq 0.2$\,au/Myr), the 5:7 MMR ($\dot{a}_N \leq 1$\,au/Myr), the 7:9 MMR ($\dot{a}_N \leq 2$\,au/Myr), as well as the 7:10 MMR ($\dot{a}_N \leq 0.01$\,au/Myr).

For the remaining cases, by comparing different MMRs of the same order under identical $\dot{a}_N$, it can be observed that MMRs located farther away require a larger $e_{min}$ values. To standardize the capture performance of MMRs at varying distances, we first recalibrated the obtained $e_{min}$. To this end, we denote the ratio of the MMR's nominal semimajor axis to that of Neptune as $\alpha=(\frac{q}{p})^{2/3}$, finding that the value of $\frac{\alpha}{\alpha-1}e_{min}$ remains roughly consistent across different positions for the same order of MMRs. The reciprocal of the coefficient $\frac{\alpha}{\alpha-1}$ also carries physical significance: when the eccentricity of a small body reaches $\frac{\alpha-1}{\alpha}$, its perihelion reaches the orbit of Neptune. For convenience, we hereafter refer to $\frac{\alpha-1}{\alpha}$ as the critical eccentricity, denoted as $e_c$.

\begin{figure}[!htb]
\centering
\resizebox{\hsize}{!}{\includegraphics{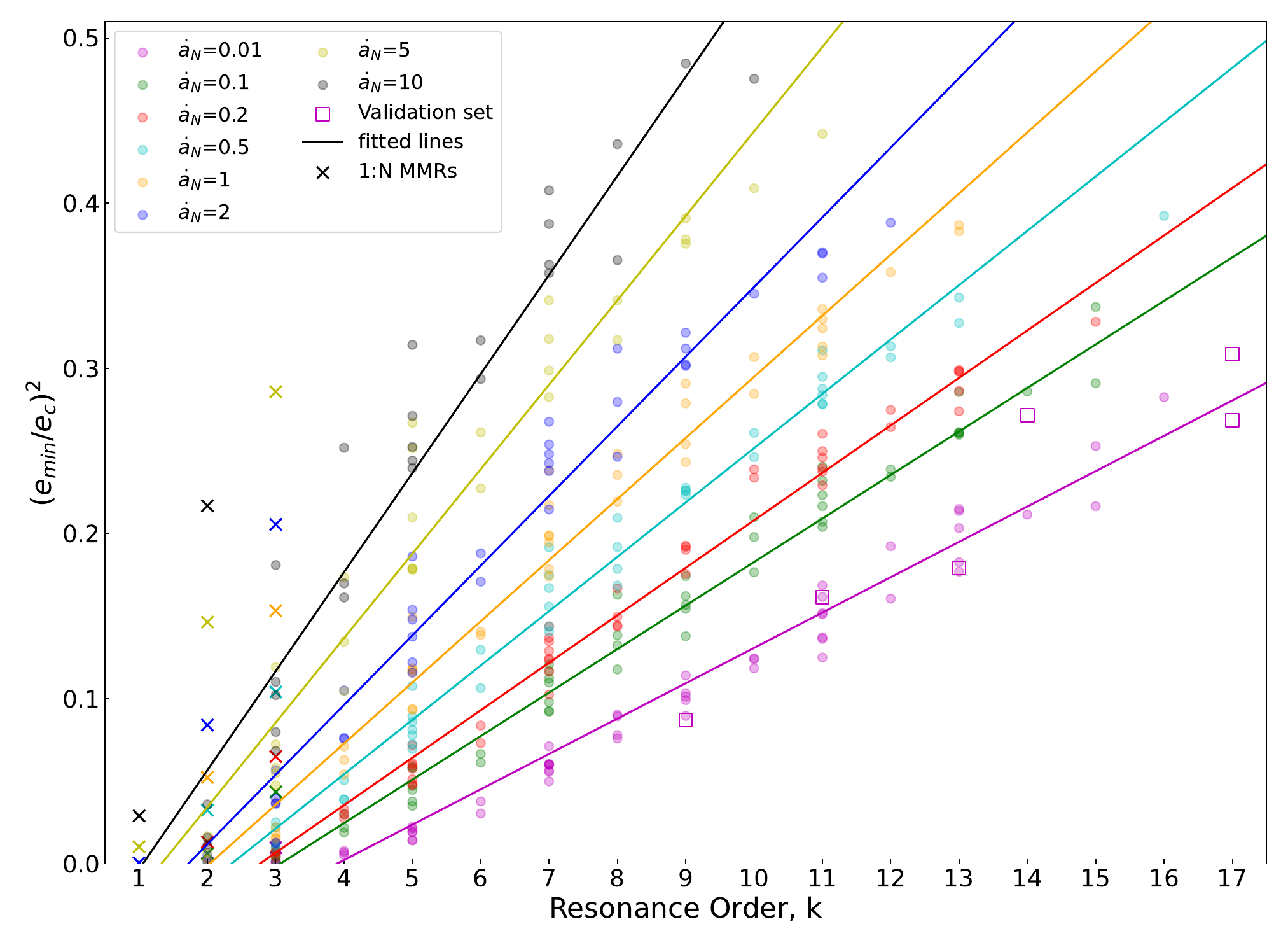}}
\caption{Relationship between $(\frac{e_{min}}{e_c})^2$ and the resonance order $k$ under different migration rates $\dot{a}_N$. The horizontal axis represents $k$, and the vertical axis represents the value of $(\frac{e{min}}{e_c})^2$. Different colors indicate different $\dot{a}_N$, and the lines represent the fitting results based on Eqs. \ref{eq1} and \ref{eq2} for the corresponding $\dot{a}_N$. Cross marks denote the 1:$q$ MMRs. The purple squares represent the validation set for the fit, and these MMRs were recorded with more than ten particles only when $\dot{a}_N=0.01$\,au/Myr.}
\label{fig:emin_fit}
\end{figure}

Following the recalibration, we examined the new quantity $\frac{e_{min}}{e_c}$. In Fig.~\ref{fig:emin_fit}, we present the relationship between $(\frac{e_{min}}{e_c})^2$ and the resonance order, $k$, under different $\dot{a}_N$. We observe that $(\frac{e_{min}}{e_c})^2$ essentially follows a linear relationship with the resonance order, $k$.

Moreover, we note that if linear fitting is applied to describe the relationship between $(\frac{e_{min}}{e_c})^2$ and $k$ at different $\dot{a}_N$, the fitted lines roughly converge at a common point. Although this property may not hold any direct physical significance, it allows us to characterize the variation of $(\frac{e_{min}}{e_c})^2$ using a simpler function. Through fitting, we express the relationship between $(\frac{e_{min}}{e_c})^2$, the resonance order, $k$, and the migration rate, $\dot{a}_N$, as follows:

\begin{equation}
    (\frac{e_{min}}{e_c})^2 \approx f(\dot{a}_N)\times(k+0.520)-0.094
    \label{eq1}
.\end{equation}

In Eq. \ref{eq1}, the function $f(\dot{a}_N)$ depends on the $\dot{a}_N$ and reflects how the slope of the aforementioned linear relationship increases with increasing $\dot{a}_N$. A subsequent fitting indicates that the $f(\dot{a}_N)$ is well approximated by an exponential function, namely,

\begin{equation}
    f(\dot{a}_N) \approx 0.020\times \dot{a}_N^{0.334}+0.017
    \label{eq2}
.\end{equation}

In Fig.~\ref{fig:emin_fit}, we also present the fitting results for different $\dot{a}_N$ values based on Eqs. \ref{eq1} and \ref{eq2}, represented by lines in corresponding colors. Despite their relatively simple forms, Eqs. \ref{eq1} and \ref{eq2} still provide a satisfactory estimation. They effectively capture the phenomenon that as the $k$ or $\dot{a}_N$ increases, the value of $(\frac{e{min}}{e_c})^2$ also rises correspondingly. The fitting remains robust even under conditions of rapid migration or high resonant orders, yielding a coefficient of determination of $R^2=0.97$.

In our simulations, six MMRs were identified with more than ten objects in the slowest simulation only, with $\dot{a}_N=0.01$\,au/Myr: 10:19, 8:25, 7:24, 5:19, 4:17, and 3:14 MMRs. With the exception of the 10:19 MMR, which is relatively close, all of them are located beyond the 1:3 MMR. These MMRs are not plotted in Fig.~\ref{fig:e_minimum} and  their $e_{min}$ values were not included in the fitting for Eqs. \ref{eq1} and \ref{eq2}. Here, we used them as a validation set, marking them with purple square symbols in Fig.~\ref{fig:emin_fit}. It can be seen that their positions still lie approximately along the fitted purple line, demonstrating the validity of Eqs. \ref{eq1} and \ref{eq2}.

However, two issues warrant attention. First, Eqs. \ref{eq1} and \ref{eq2} provide only a qualitative description of the variation pattern of $e_{min}$ and their functional forms do not carry any corresponding physical significance. In particular, the origin data used for fitting were derived from numerical simulations and represent only an approximate value, serving as an upper limit on the true $e_{min}$. Since the true $e_{min}$ should be more or less smaller, the fitting results shown in Fig.~\ref{fig:emin_fit} likely overestimate $e_{min}$ to some extent.

For example, based on Eqs. \ref{eq1} and \ref{eq2}, it can be inferred that as the migration rate $\dot{a}_N$ approaches 0, we obtain $(\frac{e{min}}{e_c})^2 \approx 0.017\times(k+0.520)-0.094$. This expression becomes nearly zero when $k=5$. Therefore, it can be deduced that no matter how slow Neptune's migration, those high-order MMRs with $k\geq6$ are incapable of capturing small bodies with $e_0=0$. However, given that this fitting serves as an upper estimate for the true $e_{min}$ (which is likely slightly lower), further research and validation are necessary to  confirm whether this conjecture is merely an artifact of the model or reflects a genuine dynamical characteristic.

Another issue is that Eqs. \ref{eq1} and \ref{eq2} exhibit significant deviation in predicting 1:$q$-type MMRs. In the above simulations, three 1:$q$ MMRs from 1:2 to 1:4 were identified, which have been marked with cross symbols in Fig.~\ref{fig:emin_fit}. However, it can be observed that all 1:$q$ MMRs consistently exhibit a higher $e_{min}$ than predicted. Therefore, we excluded these three 1:$q$ MMRs from the fitting process described earlier. The most prominent characteristic of 1:$q$ MMRs is that they are the farthest among MMRs of the same order. Previously, we introduced $e_c$ to correct for the influence of resonance distance and find that, with the same $k$ and $\dot{a}_N$, the value of $(\frac{e{min}}{e_c})^2$ is similar. Although this correction is empirical, it aligns well for cases where $p\geq 2$.

Compared to other MMRs, another major distinction of 1:$q$ MMRs lies in their intrinsic resonance structure, as their resonance islands contain two asymmetric libration centers; namely, leading and trailing centers \citepads[e.g.,][]{Message1958, frangakis1973, Beauge1994, Malhotra1996, Kotoulas2005, Voyatzis2005}. Furthermore, Neptune's migration can distort the phase-space structure of MMRs, breaking the symmetry between the leading and trailing islands \citepads[][]{Chiang2002, MurrayClay2005, Li2023}. This complex structure may be the reason why 1:$q$ MMRs require a relatively higher $e_{min}$ to capture small bodies.

Given that there are only three 1:$q$ MMRs within 80\,au, we conducted a dedicated set of simulations specifically targeting 1:$q$ MMRs to determine their $e_{min}$. In this set of simulations, we employed a relatively uniform selection method for particles to achieve more accurate results (as detailed in Section 2). We expanded our study to include the 1:9 MMR, which has recently been identified to have two resonators, suggesting a potentially large population size \citepads[][]{Volk2018, Crompvoets2022}.

When we focus specifically on 1:$q$ MMRs, parameters such as the corresponding $k$ and $\alpha$ can be directly determined once $q$ is specified. Therefore, we can adopt a simpler functional form. We eliminated the correction based on $e_c$ and directly fitted the variation pattern of $e_{min}$ using a expression similar to that given Eqs. \ref{eq1} and \ref{eq2}, while reducing one parameter (the constant term in Eq. \ref{eq2}). Through fitting, we obtain that the $e_{min}$ for 1:$q$ MMRs can be approximately expressed by the following empirical formula,

\begin{equation}
    e_{min}^2 \approx 0.043\times \dot{a}_N^{0.130}\times(k+0.804)-0.098
    \label{eq3}
.\end{equation}

\begin{figure}[!htb]
\centering
\resizebox{\hsize}{!}{\includegraphics{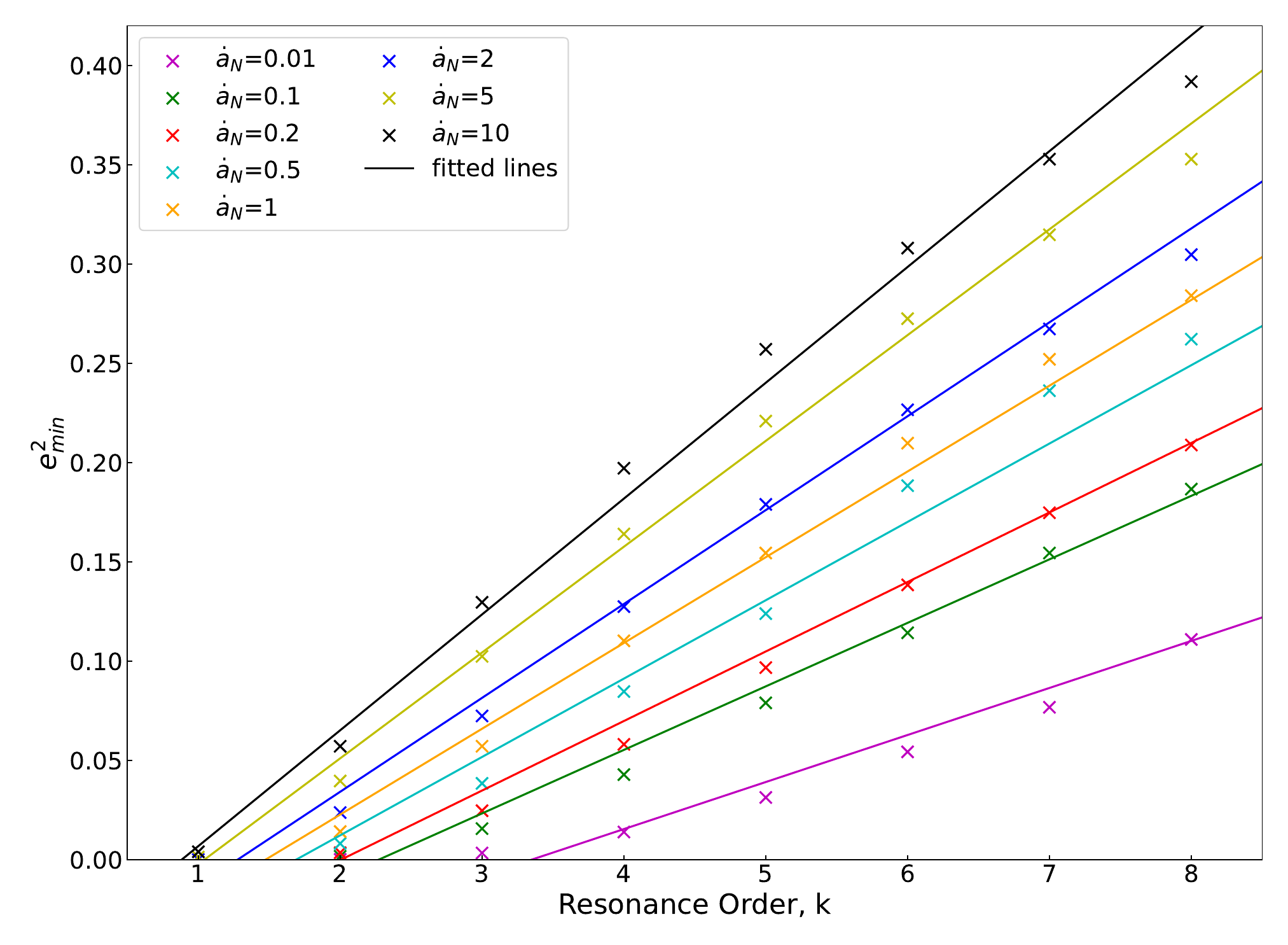}}
\caption{Relationship between $e_{min}^2$ and the resonance order $k$ under different migration rates $\dot{a}_N$ for 1:$q$ MMRs. The horizontal axis represents $k$ and the vertical axis represents the value of $e_{min}^2$. Different colors indicate various $\dot{a}_N$, and the lines represent the fitting results based on Eq. \ref{eq3} for each $\dot{a}_N$.}
\label{fig:emin_1n}
\end{figure}

In Fig.~\ref{fig:emin_1n}, we display the distribution of $e_{min}^2$ for 1:$q$ MMRs, along with the fitting results using Eq. \ref{eq3}. As a result of the more accurate and equitable determination of $e_{min}^2$ in this simulation set, the fitting performance provided by Eq. \ref{eq3} is more robust, achieving a coefficient of determination of 0.99.

After deriving Eq. \ref{eq3}, we can roughly compare its predictions with those of Eqs. \ref{eq1} and \ref{eq2}. For $q=2$, both formulations predict an $e_{min}$ close to zero. However, for cases where $q\geq3$, if the fitting results derived from non-1:$q$ MMRs are arbitrarily applied to 1:$q$ MMRs, the values given by Eqs. \ref{eq1} and \ref{eq2} are, on average, lower by 0.15 compared to the ``true values'' provided by Eq. \ref{eq3}. In other words, a relative deviation of approximately 40\%, highlighting the distinctly different behaviors of $e_{min}$ for 1:$q$ MMRs compared to non-1:$q$ MMRs.

As detailed in Section 2.1, our simulations have covered most migration rates, resonance orders, and semimajor axis ranges of practical interest. Eqs. \ref{eq1}, \ref{eq2}, and \ref{eq3} are empirical formulas obtained by fitting within this parameter space. However, it is important to note, based on their functional forms alone, that  Eqs. \ref{eq1} and \ref{eq3} can both predict $e_{min} > 1$ under very fast $\dot{a}_N$ and/or very high $k$, which is clearly unphysical. In fact, indications of this limitation can already be seen from Fig.~\ref{fig:emin_1n}: when the $e_{min}^2 \gtrsim 0.35$ ($e_{min} \gtrsim 0.6$), the data points no longer closely follow the linear trend, but mostly fall below the fitted line. This suggests that the fits provided by Eqs. \ref{eq1} and \ref{eq3} are likely more effective in the low-to-moderate eccentricity regime and require further refinement for extremely high eccentricities. In other words, caution is advised when applying these fitting formulas under conditions of extremely fast migration and high resonance orders.

\subsection{Efficiency of MMR capture}

Compared to the $e_{min}$ of MMRs, the question of how many small bodies an MMR can capture is more complex. One key consideration is that as $\dot{a}_N$ increases, $e_{min}$ gradually rises as well, which reduces the number of small bodies suitable for capture. Additionally, when Neptune approaches, if the perihelion of a small body is too close to Neptune, it might get scattered away before resonant capture can occur. Therefore, we imposed an upper eccentricity limit $e_{max}$, defined such that the perihelion of an object can reach within 32\,au (i.e., $e_{max}=1-\frac{32}{30\alpha}$). Small bodies with initial eccentricities exceeding $e_{max}$ are not counted in our statistics. Captures in this regime are rare and do not result purely from Neptune’s migration; instead, they involve either eccentricity damping through resonance crossing prior to capture or being initially placed inside the MMRs. For more distant MMRs (where $e_{max}$>0.35), the eccentricity upper limit is set by a maximum eccentricity of 0.35 defined in the simulations.

\begin{figure}[!htb]
\centering
\resizebox{\hsize}{!}{\includegraphics{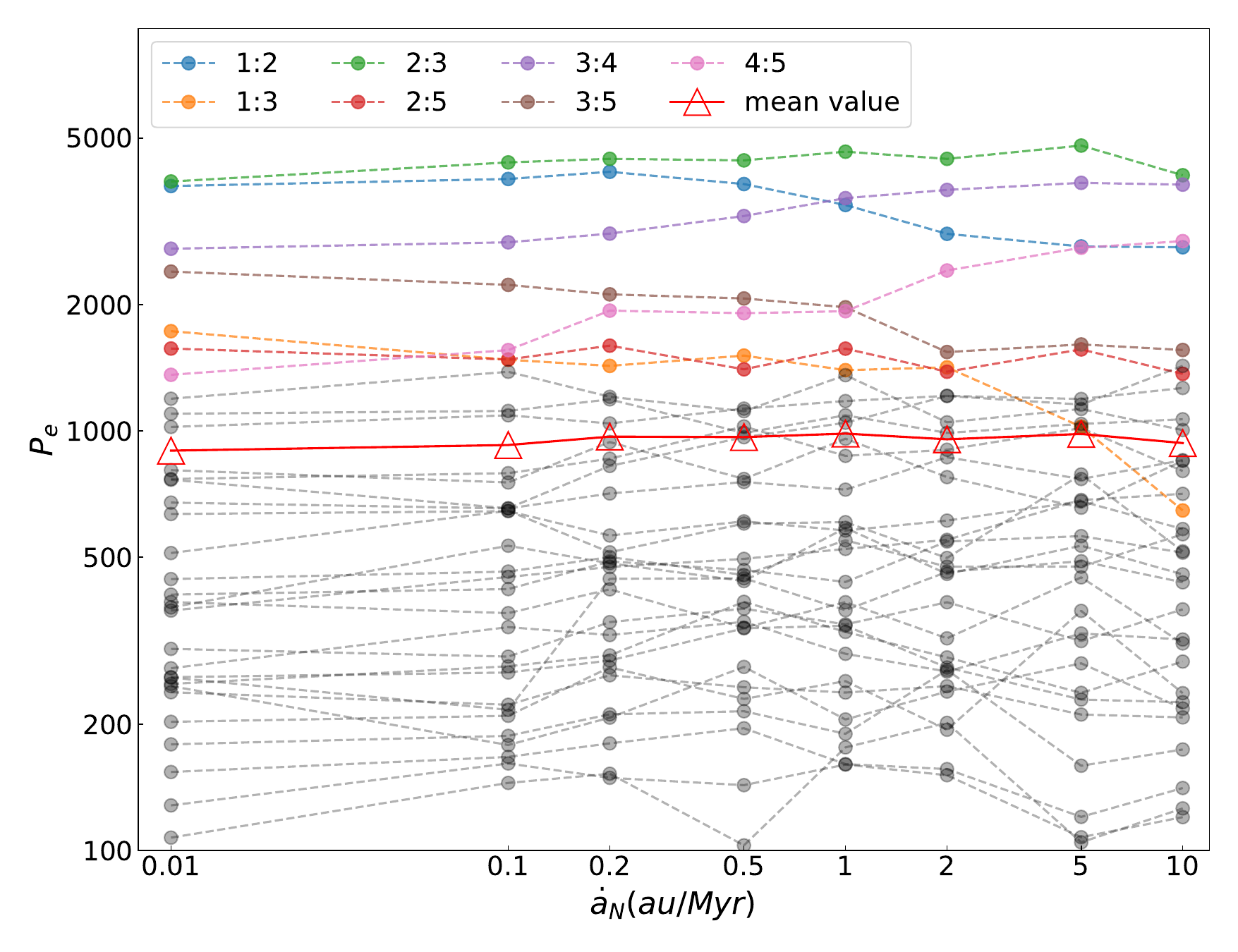}}
\caption{ Variation in the capture efficiency of MMRs under different $\dot{a}_N$.The horizontal axis represents the migration rate, $\dot{a}_N$, and the vertical axis represents the capture number per unit eccentricity, $P_e$, both on logarithmic scales. The seven MMRs with the strongest capture capability are denoted by colored lines, while the others are shown in gray. The average result across all MMRs is represented by the solid red line with triangular markers.}
\label{fig:capeff_v}
\end{figure}

To compare the capture capabilities of MMRs under different $\dot{a}_N$, we first corrected for the decrease in $P_{res}$ caused by the increase in $e_{min}$. Thus, we define the ``capture number per unit eccentricity,'' $P_e=\frac{P_{res}}{e_{max}-e_{min}}$, to describe the capture efficiency of an MMR within the eccentricity interval suitable for capture. Based on the previous subsection, we selected 33 MMRs that captured more than ten particles in all eight simulations and calculated their $P_e$, as presented in Fig.~\ref{fig:capeff_v}.

As can be seen in Fig.~\ref{fig:capeff_v}, the lines corresponding to each MMR generally exhibit a horizontal trend, indicating that after correction, $P_e$ is no longer significantly correlated with $\dot{a}_N$, but mainly related to the intrinsic capture capability of the MMR itself. The average relative standard deviation is about 13.8\%, whereas the relative standard deviation for the average capture number across all MMRs is only 3.0\%. The most efficient MMRs (e.g., the 1:2, 2:3, and 3:4 MMRs) show robust capture across a wide range of $\dot{a}_N$. The $P_{res}$ result from the 2:3 MMR provides the most direct evidence that $P_e$ and $\dot{a}_N$ are unrelated, as its $e_{min}$ is always 0, meaning the suitable eccentricity range remains unchanged. Across eight different $\dot{a}_N$ values, the 2:3 MMR captures an average of 817 objects, with a standard deviation of only 48.

Nevertheless, a certain degree of fluctuation still exists, which may primarily be attributed to randomness. As a result, this fluctuation slightly increases as the $P_e$ decreases. The deviations in the estimation of $e_{min}$ might also contribute to an increase in overall fluctuations. However, the amplitude of these fluctuations typically does not exceed 100\%, which is far less than the differences between MMRs. Furthermore, a few MMRs also exhibit specific trends; for instance, as $\dot{a}_N$ increases from 0.01 to 10\,au/Myr, the $P_e$ for the 1:3 MMR decreases from 1731 to 647, while for the 4:5 MMR, it increases from 1362 to 2838. These specific trends may be due to more complex reasons, such as the interference between different MMRs.

Next, we assume there is no significant correlation between the $P_e$ and the migration rate, or at least, that any such correlation is considerably smaller than the differences between different MMRs. Therefore, an average $P_e$, denoted as $\overline{P}_e$, can be used to characterize the capture ability of an MMR. By comparing the $\overline{P}_e$ values for different MMRs, we find that a simple exponential expression can be utilized to estimate the $\overline{P}_e$ value,

\begin{equation}
    \overline{P}_e \approx 49316\times e^{-0.475p-1.305\alpha}
    \label{eq:pe}
.\end{equation}

\begin{figure}[!htb]
\centering
\resizebox{\hsize}{!}{\includegraphics{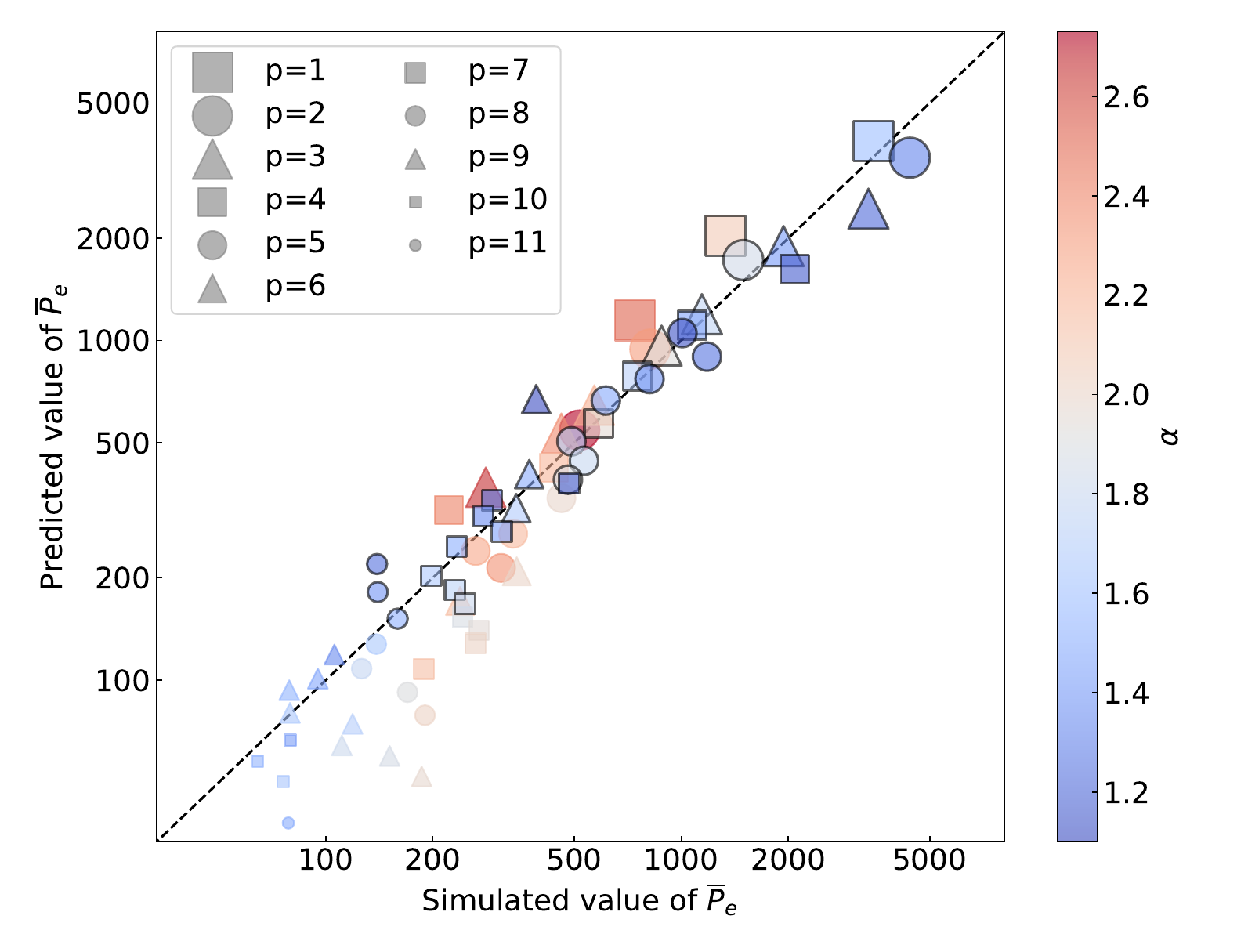}}
\caption{Distribution of the average capture capability, $\overline{P}_e$, across different MMRs. The horizontal axis represents the measured $\overline{P}_e$ from simulation data, while the vertical axis represents the $\overline{P}_e$ predicted by Eq. \ref{eq:pe}, both on logarithmic scales. MMRs with different $p$ values are denoted by different markers and the color indicates the $\alpha$ value of the MMR. Points with black borders are the MMRs that have capture records in all simulations, while those without borders are the MMRs have capture records in only some of the simulations.}
\label{fig:P_fit}
\end{figure}

The coefficients in Eq. \ref{eq:pe} are derived by fitting the data from the aforementioned 33 MMRs that captured over ten particles in all simulations. This expression indicates that the capture capability of an MMR decreases as the $p$ value increases and as the distance becomes greater. Figure~\ref{fig:P_fit} illustrates the fitting performance of Eq. \ref{eq:pe}.

Figure~\ref{fig:P_fit} displays a good fitting result with a coefficient of determination of $R^2=0.95$. As a test, we also calculated the $\overline{P}_e$ for MMRs that had capture records in only some simulations. These results are displayed  in Fig.~\ref{fig:P_fit} using markers without borders and the $\overline{P}_e$ of these MMRs can still be aptly predicted by Eq. \ref{eq:pe}. There are several MMRs with relatively small $\overline{P}_e$; the simulation values often exceed the predicted values, which is likely attributed to a selection effect inherent in the data processing method. Since these MMRs are only recorded when the capture count exceeds 10, there is a systematic overestimation of their capture capabilities. 

A particularly intriguing observation is that only the $p$ value and $\alpha$ appear directly in Eq. \ref{eq:pe}, with no direct inclusion of the $q$ value or the resonance order, $k$. In practice, during the fitting process, we also experimented with other parameter combinations, such as any two of $\alpha$, $p$, $q$, and $k$. We find that the combination of $p$ and $\alpha$ yields a significantly better fit compared to other parameter combinations. When using three parameters simultaneously (e.g., $\alpha$, $p$ and $q$), the improvement in the $R^2$ was only marginal. Therefore, after careful consideration, we conclude that using only $p$ and $\alpha$ is sufficient to effectively describe the variation in $\overline{P}_e$ across different MMRs.

A plausible explanation is that for a fixed $p$, increasing $q$ results in a higher resonance order, but simultaneously, it also expands the range swept by the resonance zone. These two effects may counterbalance each other, leading to minimal changes in capture efficiency. On the other hand, the importance of $p$ lies in the fact that MMRs with lower $p$ values are spaced farther apart from each other, reducing mutual interference.

In Eq. \ref{eq:pe}, we see that $q$ indirectly influences $\overline{P}_e$ by affecting the parameter $\alpha$. When the $p$ value is small ($p\lesssim4$), $\alpha$ increases rapidly as $q$ increases, leading to a noticeable decrease in $\overline{P}_e$ and resulting in relatively large differences in capture capability among MMRs with the same $p$ value. Conversely, when the $p$ value is relatively large ($p \gtrsim 5$), a slight increase in the $q$ value does not cause significant change in $\alpha$. In this case, the capture capabilities of different MMRs are similar, which facilitates the comparison of capture capabilities among high-$p$ MMRs. For example, the $\overline{P}_e$ values for the 5:11, 5:12, 5:13, and 5:14 MMRs are 491, 532, 480, and 460, respectively.

It should be noted that due to the application of a series of data processing methods, the metric $\overline{P}_e$ used here has undergone multiple conversions. Because the absolute number of particles that an MMR can capture also depends on the specific design of the model, involving factors such as Neptune's migration distance, and the spatial and size distributions of small bodies, the focus often lies on the ratios between their numbers \citepads[e.g.,][]{Crompvoets2022}. This pertains to the relative magnitudes of $\overline{P}_e$ across different MMRs. The coefficient 49,316 in Eq. \ref{eq:pe} is not a critically important value in itself, as its magnitude primarily depends on artificial settings, particularly the particle number used in the simulation.

In addition to its usefulness in comparisons between different MMRs, the metric $\overline{P}_e$ can also be directly applied to calculate the capture probability of a given MMR. In the planetesimal disk considered in this work, there are approximately 1000 small bodies per astronomical unit. Therefore, for every astronomical unit Neptune migrates, the region swept by an MMR contains roughly $1000\alpha$ small bodies. When extended to the unit eccentricity, the density will be approximately $\frac{1000\alpha}{0.35} \approx 2857\alpha$ objects per astronomical unit. If we further assume that all particles lie within the eccentricity range suitable for capture, then the number of particles captured per astronomical unit of Neptune's migration should be approximately $\frac{\overline{P}_e}{2.7}$ (Neptune migrates a total of 3\,au, but we only count particles captured during the first 90\% of its migration). This leads to the conclusion that for a particle with an appropriate eccentricity, the probability of being captured by that MMR is approximately $\frac{\overline{P}_e}{7714\alpha}$.

\subsection{Comparison with observations}

Thus far, we have established that the $e_{min}$ required for capture by an MMR increases as its distance becomes greater, its order increases, and the migration accelerates. The number of objects captured by an MMR first depends on the eccentricity range over which resonant capture can occur. As the $\dot{a}_N$ increases, the rise in $e_{min}$ narrows this eccentricity range, resulting in a corresponding decrease in the  capture number. After correcting for the change in capture number caused by variations in $e_{min}$, the $P_e$ of an MMR shows little correlation with the $\dot{a}_N$. Hence, it can be characterized by $\overline{P}_e$, whose value decays exponentially with increasing $p$ and $\alpha$. In this subsection, we examine several simple Neptune migration models to calculate the resonant populations captured by different MMRs, and compare them with observational estimation.

Several previous studies have provided estimates of the population sizes of MMRs, with some offering comparisons among different MMRs \citepads[e.g.,][]{Gladman2012, Adams2014, Volk2016, Crompvoets2022}. Other studies have focused on specific MMRs \citepads[e.g.,][]{Pike2015, Alexandersen2016, Volk2018, Chen2019}. Building upon these works, we summarize MMRs with available estimates of population size in Table.~\ref{tab:pop}. For certain MMRs (e.g., 2:3, 1:2, 2:5, 1:5), the various estimates available from the literature demonstrate a good degree of consistency. In Table.~\ref{tab:pop}, we have preferentially adopted the most recent estimates.

\begin{table}[htb!]
        \caption{Estimated numbers of objects in various MMRs provided by previous observational works.}
        \centering
    \begin{tabular}{ccc}
        \hline
        MMR & Nominal location\,(au) & Population $(D\gtrsim100\,km)$ \\
        \hline
        4:5 & 34.8 & $160^{+700}_{-140}$\\
        3:4 & 36.3 & $800^{+1100}_{-600}$\\
        2:3 & 39.3 & $10000^{+3600}_{-3000}$\\
        3:5 & 42.2 & $5000^{+5200}_{-3000}$\\
        4:7 & 43.6 & $3000^{+4000}_{-2000}$\\
        1:2 & 47.6 & $4400^{+1500}_{-1100}$\\
        8:17 & 49.6 & $700^{+3100}_{-700}$\\
        6:13 & 50.2 & $1700^{+4300}_{-1400}$\\
        5:11 & 50.7 & $2100^{+4900}_{-1800}$\\
        3:7 & 52.8 & $3000^{+5000}_{-2300}$\\
        5:12 & 53.8 & $2400^{+5600}_{-2000}$\\
        2:5 & 55.3 & $6600^{+4100}_{-3000}$\\
        5:13 & 56.7 & $1200^{+4800}_{-1200}$\\
        3:8 & 57.7 & $2300^{+5000}_{-2000}$\\
        4:11 & 58.9 & $3900^{+9000}_{-3400}$\\
        1:3 & 62.4 & $17000^{+11000}_{-8000}$\\
        3:10 & 66.9 & $1400^{+6000}_{-1400}$\\
        2:7 & 69.2 & $2300^{+5400}_{-1900}$\\
        4:15 & 72.4 & $2600^{+12000}_{-2500}$\\
        6:23 & 73.5 & $1800^{+7400}_{-1800}$\\
        1:4 & 75.6 & $13000^{+15000}_{-8000}$\\
        4:17 & 78.7 & $3100^{+12000}_{-3000}$\\
        8:35 & 80.2 & $2600^{+11000}_{-2500}$\\
        2:9 & 81.8 & $1100^{+6000}_{-1100}$\\
        5:24 & 85.4 & $2500^{+11000}_{-2400}$\\
        1:5 & 87.7 & $11000^{+19000}_{-8000}$\\
        4:27 & 107.1 & $5000^{+23000}_{-4800}$\\
        1:9 & 129.8 & $18000^{+39000}_{-15000}$\\
        2:23 & 152.8 & $4000^{+15000}_{-4000}$\\
        \hline
    \end{tabular}
         \tablefoot{The results for the 4:5, 3:4, 3:5, and 4:7 MMRs come from \citetads{Gladman2012}; the result for the 2:3 MMR comes from \citetads{Volk2016}; the result for the 1:2 MMR comes from \citetads{Chen2019}; and the others are from \citetads{Crompvoets2022}. The magnitude range estimated by \citetads{Gladman2012} is $H_g<9.16$, while the other studies use $H_r<8.66$. These two thresholds are roughly consistent, both corresponding to small bodies with diameters $D \gtrsim 100$\,km.}
    \label{tab:pop}
\end{table}
Overall, based on the current sample sizes of observed resonant TNOs, only a few MMRs such as the 2:3, 1:2, and 2:5 MMRs have relatively well-estimated population sizes. Most MMRs have sparse observations, providing only rough estimates in terms of order of magnitude. This means that a precise comparison between observational results and theoretical predictions is premature at this stage. Among all MMRs, the population in the 2:3 MMR is the most accurately determined. In this paper, we use this MMR as a benchmark to compare the relative population differences of other MMRs against it.

Based on the resonance capture laws derived in the previous two subsections, we can calculate the number of particles captured by a given MMR in two steps. Firstly, for a given distribution of particles, we calculate how many particles within the region swept by the MMR fall into the range suitable for capture. This calculation involves Neptune's initial and final positions, as well as the number density of particles across different eccentricities and semimajor axes. Since the aim of this study is a rough comparison, we assume a uniform semimajor axis distribution, where the surface density of the planetesimal disk is proportional to $r^{-1}$, so that the ratio among different MMRs becomes independent of the specific migration start and end points of Neptune. 

We further assume a Rayleigh distribution for eccentricity, which has been shown to be an effective description for main belt asteroids \citepads[see e.g.,][]{Plummer1916, Beck1981, Malhotra2017, Liu2023}. Under the above assumption, the number of bodies suitable for capture is proportional to $\alpha\int_{e_{min}}^{e_{max}}\frac{e}{{\sigma_e}^2}\exp{(-\frac{e^2}{{\sigma_e}^2})}de$, where $e_{min}$ is given by Eq. \ref{eq1} (for non-1:$q$ MMRs) or Eq. \ref{eq3} (for 1:$q$ MMRs), $e_{max}=1-\frac{32}{30\alpha}$, and $\sigma_e$ denotes the parameter of the Rayleigh distribution. Multiplying the number of particles by the probability of capturing a particle with an appropriate eccentricity, $\frac{\overline{P}_e}{7714\alpha}$, we can obtain the number of captured objects by a specific MMR as provided in the previous subsection,

\begin{equation}
    P_{res}\propto\overline{P}_e \times (\exp(-\frac{e_{min}^2}{2\sigma_e^2})-\exp(-\frac{e_{max}^2}{2\sigma_e^2}))
    \label{eq5}
.\end{equation}

\begin{figure*}[!htb]
\centering
\resizebox{\hsize}{!}{\includegraphics{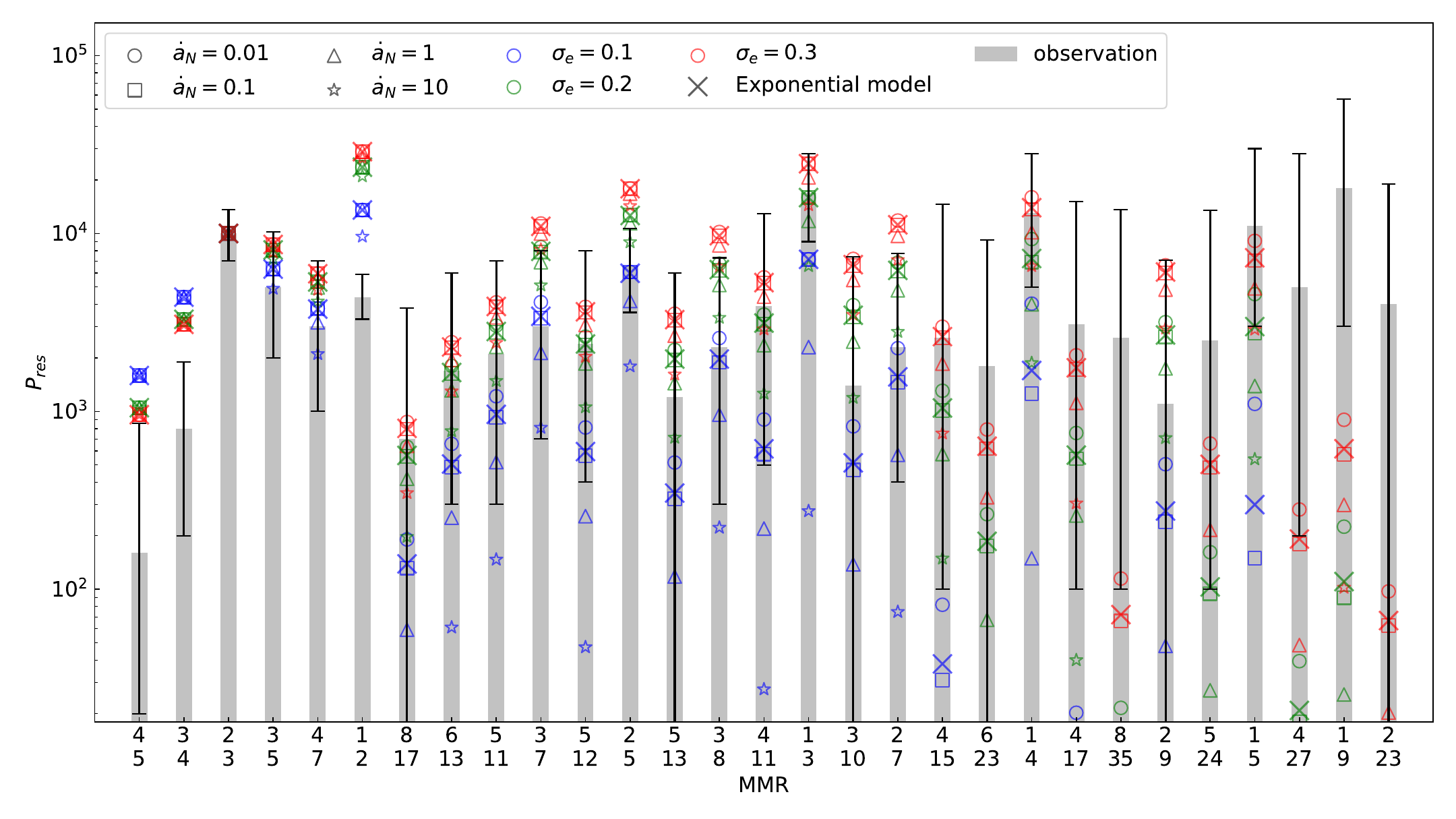}}
\caption{Captured population for each MMR under different parameters, compared with observations. The histogram displays the current observational results, with error bars indicating observational uncertainties (details in Table \ref{tab:pop}). Different markers denote different migration rates, while the crosses specifically denote the exponential migration model with an average migration rate of 0.1\,au/Myr. Different colors represent different eccentricity excitations for the small bodies. The x-axis provides the corresponding $p$ (above) and $q$ (below) values for each MMR, arranged at equal intervals in order of proximity. The y-axis represents the population within each MMR, all data are normalized, with the 2:3 MMR as a reference standard.}
\label{fig:P_pre}
\end{figure*}

Under this idealized model, the entire calculation of Eq. \ref{eq5} involves only two parameters: $\sigma_e$, which describes the level of eccentricity excitation of particles, and $\dot{a}_N$, which describes the migration rate. For this calculation, we scaled all $P_{res}$ values by a uniform factor so that $P_{res}$ for the 2:3 MMR matches the observed value of 10,000. Selecting the 2:3 MMR as a reference standard is due to its largest observed population and the most robust population estimate, as well as its superior capture ability, which remains unaffected by the migration rate (as shown by Fig.~\ref{fig:capeff_v}). We selected three different $\sigma_e$ values: 0.1, 0.2, and 0.3, corresponding to average eccentricities of 0.125, 0.251, and 0.376, respectively. Four different $\dot{a}_N$ values were tested, ranging from 0.01\,au/Myr to 10\,au/Myr. Fig.~\ref{fig:P_pre} illustrates the results obtained using these different parameters and compares them against the observational values.

Regarding the influence of the eccentricity excitation of the planetesimals, most MMRs tend to obtain a larger $P_{res}$ in a disk with a higher $\sigma_e$. This does not indicate a higher capture probability for individual planetesimals at high eccentricity, but rather reflects that the vast majority of particles in a dynamically cold disk are unsuitable for capture. However, for the two innermost MMRs, 3:4 and 4:5, this trend is reversed because they only exhibit good capture capability at lower eccentricities, compared to the 2:3 MMR. On the other hand, the effect of the migration rate is indirect: it alters the captured population by influencing $e_{min}$. This effect is particularly pronounced when $\sigma_e=0.1$, because the variation in $e_{min}$ induced by $\dot{a}_N$ can significantly alter the number of particles suitable for capture, potentially leading to an order-of-magnitude change in $P_{res}$.

For MMRs interior to the 1:2 MMR, the $P_{res}$ is relatively less sensitive to the parameters. In particular, the $P_{res}$ for the 3:4 and 4:5 MMRs are solely dependent on $\sigma_e$ and not on $\dot{a}_N$, making their observational values a crucial information. Within this region, all theoretical predictions are higher than the observations. For the nearer 3:4 and 4:5 MMRs, the population estimates still carry large uncertainties due to the limited sample sizes. However, no parameter combination yields theoretical values within the confidence interval of the observations. This discrepancy reflects a significant overestimation of $P_{res}$ for the 3:4 and 4:5 MMRs by the assumptions used in this study. On one hand, compared to the 2:3 MMR, the particles in the 3:4 and 4:5 MMRs may be significantly less stable due to their proximity to Neptune \citepads[e.g.,][]{Kotoulas2004}. On the other hand, these inner regions might have been unfavorable for planetesimal growth during the formation of the giant planets, leading to a sparser population of planetesimals.

Another MMR with a notably higher prediction is the 1:2 MMR. Current observations suggest that the population in the 1:2 MMR is only about half that of the 2:3 MMR, but simulations show that their populations are at least comparable. The primary reason for this discrepancy lies in the poorer stability of the 1:2 MMR. \citetads{Melita2000} pointed out that compared to the 2:3 MMR, the resonant islands of the 1:2 MMR are more fragmented, especially at low inclinations, which should prevent it from hosting a large population of bodies. Through frequency analysis of the 1:2 and 1:3 MMRs, \citetads{Li2024} also found that the latter exhibits a cleaner dynamical map and longer dynamical stability timescales, attributed to its greater distance and slower eigenfrequency. A similar phenomenon is observed for the 3:5 and 4:7 MMRs, whose predicted values are slightly higher than the observations (though still within the error margins). If observational errors are excluded, this might simply be due to the slightly poorer stability of these two MMRs compared to the 2:3 MMR \citepads[e.g.,][]{Melita2000}.

In the region beyond the 1:2 MMR, extending up to around the 1:4 MMR, we note that the observed populations for most MMRs in this range agree with the theoretical $P_{res}$ at least in order of magnitude and exhibit nearly identical trends. This suggests a high probability that the small bodies within these MMRs were captured by Neptune through migration, leading to several indirect inferences: Firstly, the primordial planetesimal disk potentially extended to around 60 to 70\,au, maintaining a relatively flat distribution without a sharp density drop-off. Secondly, these MMRs likely possess good long-term stability due to their greater distances, suggesting that their ability to retain objects is comparable to that of the 2:3 MMR. Thirdly, the scenario of rapid migration ($\dot{a}_N \geq 1$\,au/Myr), combined with a dynamically cold planetesimal disk ($\sigma_e=0.1$) can be ruled out, as under such conditions, the vast majority of small bodies would fail to meet the condition $e \geq e_{min}$.

The situation becomes more complex for more distant MMRs, and the most prominent issue is the further increase in observational uncertainties. In this region, theoretical predictions are generally lower than the observational estimates. This discrepancy may partly stem from randomness, as the extreme distance means that even a few sporadic detections can lead to large population estimates. Meanwhile, many other MMRs in this region, such as the 1:6 MMR, are currently lacking observational data.

Moreover, at such remote locations, the primary mechanism for supplementing resonant bodies might be the resonant sticking effect, and these existing bodies may not constitute the primordial population captured by Neptune's migration but are only temporarily trapped within these MMRs \citepads[][]{Yu2018}. \citetads{Crompvoets2022} have also noted that predictions based on resonant sticking effects align better with observations than migration models. Overall, however, explaining the existence of a large population in such distant regions remains challenging, necessitating more observational samples and higher resolution simulations in the future.

In more realistic models of the Solar System, Neptune is often considered to have undergone migration following an exponential decay pattern, a model commonly referred to as the exponential migration model \citepads[e.g.,][]{Malhotra1995}. Subsequently, it was proposed that Neptune's migration was not smooth but rather grainy \citepads{Nesvorny2016, Lawler2019}, and may even have involved a major jumping \citepads{Nesvorny2012, Nesvorny2015a}. So far, we have established a method for calculating capture into Neptune's MMRs under a constant migration rate. This method can be readily extended to scenarios where the migration rate varies. Here, we adopt a simple exponential migration model as an illustrative example.

In an exponential migration model, Neptune starts from $a_{Ni}$ and finally stops at $a_{Nf}$. The evolution of Neptune's semimajor axis can be expressed as $a_N=a_{Nf}-(a_{Nf}-a_{Ni})\exp({-\frac{t}{\tau})}$, where $\tau$ is the migration timescale. The migration rate $\dot{a}_N$ can then be written as a function of semimajor axis, namely $\dot{a}_N=\frac{a_{Nf}-a_{N}}{\tau}$. Therefore, the average migration rate is easily obtained as $\frac{a_{Nf}-a_{Ni}}{2\tau}$. It should be noted that this average is taken over different semimajor axes, not over time. On the other hand, the captured number by a specific MMR can be obtained by integrating over different semimajor axes, namely,

\begin{equation}
    P_{res}\propto\overline{P}_e \int_{a_{Ni}}^{a_{Nf}} (\exp(-\frac{e_{min}^2}{2\sigma_e^2})-\exp(-\frac{e_{max}^2}{2\sigma_e^2}))d a_{N}
    \label{eq6}
.\end{equation}

Generally, Eq. \ref{eq6} would also need to account for the possibility that the planetesimal density and $\sigma_e$ are functions of semimajor axis. However, under our assumptions, the planetesimals follow a uniform semimajor axis distribution and a consistent eccentricity excitation, and both $e_{max}$ and $P_e$ in Eq. \ref{eq6} are also unaffected by $\dot{a}_N$. Consequently, the only quantity in Eq. \ref{eq6} that depends on $a_N$ is $e_{min}$.

As a test case, we assumed $a_{Ni} = 25$\,au, $a_{Nf} = 30$\,au, and $\tau = 25$\,Myr. Under different $\sigma_e$, we calculated the capture capability of various MMRs within this exponential migration model and marked them with cross symbols in Fig.~\ref{fig:P_pre}. Since the average migration rate for this model is 0.1\,au/Myr, we can conveniently compare this exponential migration model with the linear migration model for $\dot{a}_N = 0.1$\,au/Myr.

As shown in Fig.~\ref{fig:P_pre}, the results from all cross symbols and square symbols are nearly consistent. Examining cases where $P_{res} > 1$, the maximum difference introduced by the exponential model occurs for the 1:5 MMR and $\sigma_e = 0.1$, where the $P_{res}$ obtained from the exponential migration is twice that of the constant migration with $\dot{a}_N = 0.1$\,au/Myr. This difference arises because 1:$q$ MMRs have a larger capture efficiency, but also a relatively higher $e_{min}$. Consequently, the slow phase at the end of exponential migration makes a substantial contribution to their total captured population. Aside from this particular case, the average difference between the exponential and constant migration models is only 4.3\%. This minor difference is nearly indistinguishable when plotted on the logarithmic scale of Fig.~\ref{fig:P_pre}. Particularly given that current comparisons with observations are only at the order-of-magnitude level, exponential and constant migration models exhibit virtually no difference when their average migration rates are the same.

\section{Conclusion and discussion}

In this paper, we systematically investigate the ability of Neptune's exterior MMRs to capture objects and provide a detailed numerical assessment. Based on previous theories of resonance capture \citepads[e.g.,][]{Mustill2011}, a specific MMR under a given migration rate can only capture small bodies with eccentricities exceeding a certain threshold, termed the minimum capture eccentricity, $e_{min}$. Depending on the capture efficiency of each MMR, the number of captured small bodies $P_{res}$ varies. We first employed a simple model to identify particles captured by MMRs (as shown in Fig.~\ref{fig:Res_identify} and Fig.~\ref{fig:ae_final}) and determined their corresponding $e_{min}$ and $P_{res}$ under different migration rates (Fig.~\ref{fig:e_minimum}).

Then we fit $e_{min}$ and $P_{res}$ using simple functional expressions. Following the appropriate scaling, the $(\frac{e_{min}}{e_c})^2$ of MMRs roughly follows a linear relationship with their resonance order and the slope of this line can be expressed as a function of Neptune's migration rate $\dot{a}_N$ (Eqs. \ref{eq1} and \ref{eq2} and Fig.~\ref{fig:emin_fit}). Meanwhile, the pattern followed by 1:$q$ MMRs shows some differences, their $e_{min}$ are significantly higher than those of non-1:$q$ MMRs (Eq. \ref{eq3} and Fig.~\ref{fig:emin_1n}). This suggests that the unique dynamical structure of 1:$q$ MMRs may differentiate them from other MMRs, making them capture only small bodies with relatively higher eccentricities.

As for $P_{res}$, the primary conclusion is that $\dot{a}_N$ directly influences only $e_{min}$, thereby indirectly affecting $P_{res}$, while the capture efficiency per unit eccentricity, $P_e$, is almost unaffected by $\dot{a}_N$ (Fig.~\ref{fig:capeff_v}). This character leads us to use $\overline{P}_e$ to describe the probability of capturing objects. Through fitting, we find that $\overline{P}_e$ can also be expressed with a simple formula: it decays exponentially with increasing $p$ value and $\alpha$ value of the MMR (Eq. \ref{eq:pe} and Fig.~\ref{fig:P_fit}).

In this step, we made a major assumption that MMRs capture objects with roughly uniform probability within the suitable eccentricity range (from $e_{min}$ to $e_{max}$), whereas in reality this is not the case. Previous works have found that at a given migration rate, the probability of an MMR capturing a small body transitions from near certainty to near impossibility as eccentricity increases. Furthermore, for second-order MMRs, this transition occurs over a broader eccentricity range compared to first-order MMRs. \citepads[e.g.,][]{Mustill2011}. Therefore, it is reasonable to speculate that the assumption of the capture probability being independent of eccentricity may only be approximately valid for higher order MMRs and not for lower order MMRs. However, this assumption does not impact the numerical conclusions drawn in this study.

One of the main objectives of this paper is to compare the capture capabilities among different MMRs through numerical evaluation, with the two metrics, $e_{min}$ and $\overline{P}_e$, serving as valuable aids. Generally, a ``stronger'' MMR will be able to capture at lower $e_{min}$ values, while also exhibiting a higher $\overline{P}_e$. However, these two metrics define the strength of MMRs in slightly different ways. According to Eqs. \ref{eq1} and \ref{eq3}, MMRs that are closer (having smaller $e_c$), at a lower order, and non-1:$q$ type tend to have lower $e_{min}$ values. While Eq. \ref{eq:pe} indicates that MMRs that are closer and have smaller $p$ values exhibit higher capture efficiency. From this perspective, the capture efficiency of 1:$q$ MMRs is significantly better than that of neighboring MMRs. Therefore, in any sense, a closer MMR is stronger, but $e_{min}$ and $\overline{P}_e$ tend to favor lower orders and smaller $p$ values, respectively; thus, comparisons of $e_{min}$ and $\overline{P}_e$ between two MMRs lead to opposite conclusions in some special cases. Taking a set of third-order MMRs as an example, if ranked by $e_{min}$ from strongest to weakest, the order would be 7:10 > 5:8 > 4:7 > 2:5, because of the increasing distance. Meanwhile, if they are ranked by $\overline{P}_e$, the order becomes 7:10 < 5:8 < 4:7 < 2:5, as a result of decreasing $p$ values.

Previous studies have proposed various metrics to characterize the strength of MMRs. For instance, \citetads{Gallardo2006} defined the resonance strength as the difference between the average and minimum values of the disturbing function. When studying resonant sticking, the strength can be measured by the probability of a particle being transiently captured by the MMRs or the timescale it remains trapped \citepads[e.g.,][]{Lykawka2007,Yu2018}. Additionally, \citetads{Lan2019} quantified the resonant area for MMRs by measuring the widths of their resonant zones. These metrics share a common feature that 1:$q$ type MMRs dominate the region beyond Neptune, followed by 2:$q$ type MMRs, which is similar to the pattern observed with the metric $\overline{P}_e$.

However, there are nuanced differences among these metrics. For instance, under a fixed $p$ value, for resonant sticking effects, more distant MMRs tend to have better stability timescales due to their longer orbital periods and greater distance from planetary perturbations \citepads[e.g.,][]{Yu2018}. The resonant areas show an increasing trend with distance for closer in MMRs (with the exception of $p$=2, where the 2:3 MMR has a larger area than the 2:5 MMR) \citepads{Lan2019}. Although the resonant area decreases at greater distances, the decline is gradual, so distant MMRs still possess considerable areas. Whereas the $\overline{P}_e$ exhibits an exponential decrease with distance. Regarding the ``resonance strength'' calculated by \citetads{Gallardo2006}, aside from $p$ values, resonance orders also contribute a great deal. Furthermore, closer, low-order MMRs exhibit high resonance strength, making this metric somewhat similar to $e_{min}$ in that sense. Therefore, in different contexts, it is essential to carefully select the appropriate metric for comparing MMRs to achieve optimal results.

We then applied the aforementioned method for calculating captured populations under various parameters and compared the results with observations (Table \ref{tab:pop} and Fig.~\ref{fig:P_pre}). Taking 2:3 MMR as a reference, the theoretical capture abilities of MMRs interior to the 1:2 MMR all exceed their observed populations more or less, which likely reflects the superior long-term stability of the 2:3 MMR compared to these MMRs. This helps improve our understanding of the density distribution of the primordial planetesimal disk and the long-term structure and stability. In the region between the 1:2 and 1:4 MMRs, the theoretical estimates of captures align well with observational values. This suggests the possible significant contribution of migration capture mechanisms to the populations of MMRs in this region, providing valuable insights into the details of the dynamical evolution of the early Solar System. Beyond the 1:4 MMR, the theoretical prediction is generally much lower than observational estimates. This indicates that the populations in these distant MMRs require replenishment through other mechanisms, such as resonant sticking \citepads[e.g.,][]{Lykawka2007,Yu2018}. 

This method for calculating the expected number of captures can also be readily extended to exponential migration models. Given the same average $\dot{a}_N$, exponential and constant migration models exhibit only minor differences. For 1:$q$ MMRs, however, these differences are somewhat more pronounced. This is because such MMRs have relatively high $\overline{P}_e$ yet require higher $e_{min}$, and the slow-phase of exponential migration contributes significantly to their capture numbers. Given the population sizes of individual MMRs can be determined with greater precision observationally, this characteristic could serve as a key to reconstructing the details of migration in the future.

To better understand the early evolution of the Solar System, there are several directions for future research efforts. First, in this study, we primarily considered a planar model without incorporating the influence of orbital inclinations, which could alter the required $e_{min}$ for capture. Previous studies have explored the impact of orbital inclinations \citepads[e.g.,][]{Li2014, Namouni2015, Namouni2017}. A more accurate assessment of the capture ability of Neptune's MMRs within a full 3D framework awaits more systematic investigation.

Interference between neighboring MMRs was also not considered. When an MMR sweeps through a region, particles in that area might have already been carried away by the previously sweeping MMRs or their orbits may have been modified even if they were not captured \citepads[][]{Murray1999, Mustill2011}. In this paper, we  endeavor to minimize such interference. Specifically, we employed a continuously distributed planetesimal disk and limited Neptune's migration to a relatively short distance, ensuring that the regions covered by the main MMRs do not overlap with each other. When studying 1:$q$ MMRs, such an interference is exceptionally small because, in these simulations, Neptune migrated only 1\,au and there are no other major MMRs in the vicinity of a 1:$q$ MMR. For densely packed high-order MMRs, mutual interference is unavoidable. Nevertheless, due to their lower capture efficiency and limited ability to modify orbits, the interference among these weaker MMRs is negligible. In the real Solar System, such mutual interference inevitably occurs, especially for the innermost few first-order MMRs (from the 3:4 to the 6:7 MMRs). This is where the interference could be significant because they all have the ability to efficiently capture particles at low eccentricities. Moreover, such interference from the 2:3 MMR might be one of the reasons why there are few particles observed in the 3:4 and 4:5 MMRs.

Another crucial issue lies in the long-term stability of various MMRs. While qualitative conclusions have been drawn for specific MMRs in the past, a unified quantitative comparison is lacking. For instance, as previously mentioned, the observed numbers of bodies in the 3:4 and 4:5 MMRs are much lower than expected, which could be due to their poor long-term stability or because these closer regions are inherently unfavorable for planetesimal growth. In this study, there is some indirect speculation about the long-term stability of various MMRs, but direct numerical simulations will be a key component in future research.

Lastly, and most importantly, the current observational data on TNOs remain relatively scarce, particularly for resonant TNOs. Most MMRs have only a handful of detected bodies, which not only fails to eliminate the possibility of random noise, but also hinders the precise estimation of their population. Consequently, comparisons between observations and theoretical predictions are often limited to an order of magnitude level. With the advancement of observational instruments and the gradual expansion of surveys, we anticipate the discovery of more resonant TNOs, which will provide richer information for unraveling the history of planetary migration and offer insights into the dynamics of the early Solar System.

\begin{acknowledgements}
        This work is supported by the Science and Technology Development Fund (FDCT) of Macau (grant Nos. 0034/2024/AMJ, 0008/2024/AKP, 002/2024/SKL, 0002/2025/AKP) and the National Key Research and Development Program of China (grant No.2024YFE0201000). Li-Yong Zhou thanks the support from National Natural Science Foundation of China (NSFC, Grants No.12373081 \& No.12150009) and the China Manned Space Program with grant No.CMS-CSST-2025-A16.
\end{acknowledgements}

\bibliographystyle{aa-note}
\bibliography{ResCap}
\end{document}